\documentclass[nofootinbib,aps,11pt]{revtex4}
\usepackage{graphicx}
\usepackage{cancel}
\usepackage{amssymb}
\usepackage{textcomp}
\usepackage{amsmath}
\usepackage{bm}
\usepackage{times}
\usepackage{epsfig}
\usepackage{color}
\usepackage{citesort}

\begin{document}
\preprint{MPP-2008-85}
\vfill
\preprint{}
\vspace{2.0cm}
\title{\Large Baryogenesis via Leptogenesis in Adjoint SU(5)}
\vspace{4.0cm}
\author{Steve Blanchet$^{1}$, Pavel Fileviez P\'erez$^{2}$}
\affiliation{
$^{1}$ Max Planck Institute for Physics (Werner-Heisenberg-Institute) \\
Foehringer Ring 6, 80805 Munich, Germany 
\\
$^{2}$ University of Wisconsin-Madison, Department of Physics\\
1150 University Avenue, Madison, WI 53706, USA}
\date{\today}
\begin{abstract}
The possibility to explain the baryon asymmetry in the Universe
through the leptogenesis mechanism in the context of Adjoint $SU(5)$
is investigated. In this model neutrino masses are generated
through the Type I and Type III seesaw mechanisms, and the
field responsible for the Type III seesaw, called $\rho_3$,
generates the $B-L$ asymmetry needed to satisfy the observed value of 
the baryon asymmetry in the Universe. We find that 
the $C\!P$ asymmetry originates only from the vertex correction, since
the self-energy contribution is not present. When neutrino masses
have a normal hierarchy, successful leptogenesis is possible
for $10^{11} \ \text{GeV} \lesssim M_{\rho_3}^{\rm NH} \lesssim 4\times 10^{14} \ \text{GeV}$.
When the neutrino hierarchy is inverted, the allowed mass range
changes to $2\times 10^{11} \ \text{GeV} \lesssim M_{\rho_3}^{\rm IH}
\lesssim 5\times 10^{11} \ \text{GeV}$. These constraints make possible
to rule out a large part of the parameter space in the theory which was
allowed by the unification of gauge interactions and the constraints
coming from proton decay.
\end{abstract}
\pacs{fadafad}
\maketitle
\section{Introduction}
The origin of the baryon asymmetry in the Universe is one of
the most interesting issues in modern cosmology. One of the simplest
explanations is provided by the Baryogenesis via Leptogenesis
scenario~\cite{Fukugita}, where the lepton asymmetry generated
in the out-of-equilibrium decays of the fields responsible
for the seesaw mechanism of neutrino masses~\cite{TypeI} is
converted into a baryon asymmetry by sphaleron transitions~\cite{kuzmin}.
This idea is very appealing due to the strong connection with the
origin of neutrino masses.

Grand Unified Theories (GUTs)~\cite{GUTs} are one of the most
appealing extensions of the Standard Model (SM), explaining e.g.
the origin of the SM gauge interactions and neutrino masses.
In particular, these theories provide a natural framework for the
implementation of the Baryogenesis via Leptogenesis mechanism
mentioned above. In this article we study the leptogenesis mechanism
and its possible predictions in the context of the Adjoint $SU(5)$~\cite{adjoint}.
In this theory the Higgs sector is composed of $\bm{5_H}$, $\bm{24_H}$ and
$\bm{45_H}$ Higgses and the matter lives in the $\bm{\bar{5}}$,
$\bm{10}$, and $\bm{24}$ representations. All fermion masses are
generated at the renormalizable level, and neutrino masses are generated
through the Type I and Type III~\cite{TypeIII} seesaw mechanisms.
The predictions coming from the unification of
gauge interactions and proton decay were studied in great detail
in~\cite{Hoernisa}, where the authors concluded that the
lightest fermionic field living in the adjoint representation
has to be the field responsible for the Type III seesaw mechanism.
This result strongly motivates us to study the leptogenesis mechanism
in the context of Adjoint $SU(5)$ since in this case there is no
ambiguity about which field is
responsible for leptogenesis.

Our first finding is that the $C\!P$ asymmetry is generated only
by the vertex correction since the self-energy contribution vanishes.
When the neutrino mass hierarchy is normal, successful leptogenesis
is possible in a large region of the parameter space. On the contrary,
when the spectrum for neutrinos is inverted, we find consistent solutions
for a very restricted mass range. Finally, we show that imposing the
constraints coming from leptogenesis one can rule out a large region
in the parameter space of the theory which was allowed by the unification
of gauge interactions and the constraints coming from proton decay.

The paper is organized as follows: In Section II we discuss
the theory of Adjoint $SU(5)$ and its predictions for neutrino
masses. In Section III we present our computation of
the baryon asymmetry through leptogenesis, and derive bounds
on the mass of the field responsible for the
Type III seesaw. In Section IV we discuss
the possible constraints on the spectrum of the theory from
successful leptogenesis. In the last section we summarize
our main results.
\section{Adjoint $SU(5)$ Unification and Neutrino Masses}
In the context of Adjoint $SU(5)$~\cite{adjoint} neutrino masses
are generated through the Type I~\cite{TypeI} and Type III~\cite{TypeIII}
seesaw mechanisms. In this theory the Higgs sector is composed
of $\bm{5_H}$, $\bm{24_H}$, and $\bm{45_H}$ and the matter fields live
in ${\bf{\bar{5}}}=(d^C, l)_L$, ${\bf{10}}=(u^C, Q, e^C)_L$ and
$
{\bf {24}}= (\rho_8,\rho_3, \rho_{(3,2)}, \rho_{(\bar{3}, 2)},
\rho_{0})_L=({8},{1},0)\bigoplus({1},{3},0)\bigoplus({3},{2},-5/6)
\bigoplus(\overline{{3}},{2},5/6)\bigoplus({1},{1},0)
$. In our notation $\rho_3$ and $\rho_0$ are the $SU(2)_L$
triplet responsible for Type III seesaw and the singlet
responsible for Type I seesaw, respectively. See
reference~\cite{SUSY} for the supersymmetric
version of the theory.

The relevant interactions for neutrino masses in this context
are given by:
\begin{eqnarray}
\label{neutpot}
V &=& c_{\alpha} \ \bm{\bar{5}_{\alpha} \ 24 \ 5_H} \ + \ p_{\alpha} \ \bm{\bar{5}_{\alpha} \ 24 \ 45_H}
\ + \ M \ \text{Tr} \ \bm{24}^2 \ + \ \lambda \ \text{Tr} \ \left( \bm{24}^2 \bm{24_H} \right) \ + \ \text{h.c.} 
\nonumber\\
&=& - c_{\alpha} \ \ell_\alpha^T \ {\rm i} \sigma_2 \rho_3 \ H_1 \ + \ 3 p_\alpha \ \ell_\alpha^T \ {\rm i} 
\sigma_2 \rho_3 \ H_2 \ + \ \frac{3 c_\alpha}{2 \sqrt{15}} \ \ell_\alpha^T \ {\rm i} \sigma_2 \rho_0 \ H_2 \ + \ 
\frac{\sqrt{15}}{2} p_\alpha \ \ell_\alpha^T  \ {\rm i} \sigma_2 \rho_0 \ H_2 \ + \ \ldots  
\end{eqnarray}
where $\alpha=1,2,3$, and $H_1$ and $H_2$ are the Higgs doublets living in ${\bm{5_H}}$ and ${\bm{45_H}}$, 
respectively. Once $\bm{24_H}$ gets the expectation value, $\langle \bm{24_H} \rangle = v\, {\rm diag}
(2,2,2, -3, -3) / \sqrt{30}$, the masses of the fields responsible for
seesaw living in $\bm{24}$ are given by:
\begin{eqnarray}
M_{\rho_0} &=& \left| m \ - \ \frac{\tilde{\lambda} M_{\rm GUT}}{\sqrt{\alpha_{\rm GUT}}}\right|,\ \qquad \ \text{and} \ \qquad
M_{\rho_3} = \left| m \ - \ \frac{3 \tilde{\lambda} M_{\rm GUT}}{\sqrt{\alpha_{\rm GUT}}} \right|,
\end{eqnarray}
where we have used the relations $M_V= v \sqrt{5 \pi \alpha_{\rm GUT}/3}$,
$\tilde{\lambda}= \lambda / {\sqrt{50 \pi}}$ and chose $M_V$ as the
unification scale. The predictions coming from the unification
of gauge interactions and proton stability was studied in detail
in~\cite{Hoernisa}. In this recent study the authors concluded
that in order to satisfy the unification and proton decay
constraints the field $\rho_3$ has to be the lightest field in the
$\bm{24}$ representation. Therefore, the theory~\cite{adjoint} is a good
theory for leptogenesis since one can predict which field
generates the lepton asymmetry.

Once the GUT symmetry is broken by the vev of $\bm{24_H}$, 
all elements of the mass term for the Higgs doublets are large, 
of order the GUT scale. Diagonalizing the mass matrix, one obtains one light eigenstate, 
to be identified with the SM Higgs $H$, and one heavy eigenstate $H'$
with a mass at the 
GUT scale. In particular, it is relevant for our study that 
$\rho_3$ is only kinematically allowed to decay into the SM
Higgs $H$.

Writing $H_1= \cos \alpha \ H \ - \ \sin \alpha \ H'$ and 
$H_2 = \sin \alpha \ H \ + \ \cos \alpha \ H'$, and since only $H$ gets a vev 
$\langle H \rangle=v_0/\sqrt{2}$, we have that 
$\cos \alpha = v_5/v_0$ and $\sin \alpha = v_{45}/ v_0$. The relevant terms in Eq.~(\ref{neutpot})
with the addition of the mass terms are then given by 
\begin{equation}
V_{\nu} =  h_{\alpha 1} \ \ell_{\alpha}^T \ {\rm i} \sigma_2 \ C \ \rho_3 \ H \ + \
 h_{\alpha 2} \ \ell_{\alpha}^T \ {\rm i} \sigma_2 \  C \ \rho_0 \ H \ + \
 M_{\rho_3} \ \text{Tr} \ \rho_3^T \ C \ \rho_3
\ + \ \frac{1}{2} \ M_{\rho_0} \ \rho_0^T \ C \ \rho_0 \ + \ \text{h.c.}
\end{equation}
where\footnote{Compared to \cite{adjoint} we use the more convenient definition
in the context of leptogenesis $a_{\alpha}\equiv h_{\alpha 1}$ and $b_{\alpha}
\equiv h_{\alpha 2}$, which
makes apparent the similarity between the model under consideration and the
type I seesaw model with two right-handed neutrinos.}
\begin{equation}
h_{\alpha 1} ={1\over 2 \sqrt{2} v_0} \left( c_{\alpha}  \ v_5 \ - \  3 p_{\alpha} \ v_{45} \right)
\qquad
\text{and}
\qquad
h_{\alpha 2}  = {\sqrt{15}\over 2 \sqrt{2} v_0} \left( \frac{c_{\alpha}  \ v_5}{5} \ + \ p_{\alpha} \ v_{45} \right),
\end{equation}
with $v_0=174~{\rm GeV}$. In the above equations $v_5/\sqrt{2}=\langle \bm{5_H} \rangle$,
$v_{45}/\sqrt{2} = \langle \bm{45_H} \rangle^{15}_1 = \langle \bm{45_H} \rangle^{25}_2 =
\langle \bm{45_H} \rangle^{35}_3$, and the matrix representation for $\rho_3$ is given by
\begin{equation}\label{rho3}
\rho_3 = \frac{1}{2} \left( \begin{array} {cc}
 T^0  &  \sqrt{2} T^+ \\
 \sqrt{2} T^-  & - T^0
\end{array} \right).
\end{equation}

Integrating out the fields responsible 
for the seesaw mechanism, the mass matrix for neutrinos reads as
\begin{eqnarray}
M^\nu_{\alpha \beta} & = & \left( \frac{h_{\alpha 1} \ h_{\beta 1}}{M_{\rho_3}} \ +
\ \frac{h_{\alpha 2} \ h_{\beta 2}}
{M_{\rho_0}} \right)v_0^2.
\end{eqnarray}
The theory~\cite{adjoint} predicts one massless neutrino at tree level.
Therefore, we could have either a normal neutrino mass hierarchy:
$m_1=0$, $m_2=\sqrt{\Delta m_{\rm sol}^2}$
and $m_3=\sqrt{\Delta m_{\rm sol}^2 + \Delta m_{\rm atm}^2}$, or the inverted neutrino
mass hierarchy: $m_3=0$, $m_2=\sqrt{\Delta m_{\rm atm}^2}$ and
$m_1=\sqrt{\Delta m_{\rm atm}^2 - \Delta m_{\rm sol}^2}$. $\Delta m_{\rm sol}^2
\simeq 8 \times 10^{-5}$ eV$^{2}$
and $\Delta m_{\rm atm}^2 \simeq 2.5 \times 10^{-3}$ eV$^{2}$ are the mass-squared
differences of solar and atmospheric neutrino oscillations, respectively~\cite{review}.

Finally, the kinetic terms for the fields responsible for seesaw read as
\begin{eqnarray} \label{scattint}
{\cal{L}}_{\rm kin} & = & {\rm i}\,\text{Tr} \ \bar{\rho}_3 \ \gamma^\mu \ D_\mu \ \rho_3 \ +
{\rm i}\, \bar{\rho}_0 \ \gamma^\mu \ \partial_\mu \ \rho_0
\end{eqnarray}
where $D_\mu \ \rho_3 = \partial_\mu \rho_3 \ + \ {\rm i} g_2 \ [W_\mu, \rho_3]$ and
\begin{equation}
W_\mu = \frac{1}{2} \left( \begin{array} {cc}
 W_\mu^3  &  \sqrt{2} W^+_\mu \\
 \sqrt{2} W^-_\mu  & - W_\mu^3
\end{array} \right).
\end{equation}
The gauge-scattering term coming from the kinetic term will be crucial in the
leptogenesis analysis, which we will be dealing with in the next section. A
convenient parametrization
of the $3\times 2$ Yukawa coupling matrix $h$ is the so-called Casas-Ibarra~\cite{casas}
\begin{equation}\label{ci}
h=U D_{\nu}^{1/2}\Omega D_{\rho}^{1/2}/v_0,
\end{equation}
where $U$ is the PMNS lepton mixing matrix, $D_{\nu}={\rm diag}(m_{1},m_2, m_3)$
and $D_{\rho}={\rm diag}(M_{\rho_3},M_{\rho_0})$. The $\Omega$ matrix takes here the well-known
form corresponding to the Type I seesaw case with 2 right-handed neutrinos \cite{ir}:
\begin{equation}\label{Omega}
\Omega^{\rm NH}=\left(
\begin{array}{cc}
  0  &  0   \\
  \pm \sqrt{1-{\omega}^2} & - {\omega} \\
  \xi \, {\omega} & \pm \xi\, \sqrt{1-{\omega}^2}
\end{array}
\right) ,\quad \Omega^{\rm IH}=\left(
\begin{array}{cc}
    \pm \sqrt{1-{\omega}^2 } & - {\omega}  \\
  \xi\, {\omega}  & \pm \xi\, \sqrt{1-{\omega}^2 }\\
0&0
\end{array}
\right),
\end{equation}
in the normal and inverted hierarchy, respectively, and where $\omega$ is a complex
parameter. $\xi=\pm 1$ is a discrete parameter that accounts for a discrete
indeterminacy in $\Omega$. For the PMNS matrix
$U$, we adopt the usual parametrization~\cite{PDG}
\begin{equation}\label{Umatrix}
U=\left( \begin{array}{ccc}
c_{12}\,c_{13} & s_{12}\,c_{13} & s_{13}\,{\rm e}^{-{\rm i}\,\delta} \\
-s_{12}\,c_{23}-c_{12}\,s_{23}\,s_{13}\,{\rm e}^{{\rm i}\,\delta} &
c_{12}\,c_{23}-s_{12}\,s_{23}\,s_{13}\,{\rm e}^{{\rm i}\,\delta} & s_{23}\,c_{13} \\
s_{12}\,s_{23}-c_{12}\,c_{23}\,s_{13}\,{\rm e}^{{\rm i}\,\delta}
& -c_{12}\,s_{23}-s_{12}\,c_{23}\,s_{13}\,{\rm e}^{{\rm i}\,\delta}  &
c_{23}\,c_{13}
\end{array}\right)
\times {\rm diag}(1, {\rm e}^{{\rm i}\,{\Phi/ 2}}, 1)
\, ,
\end{equation}
where $s_{ij}\equiv \sin\theta_{ij}$,
$c_{ij}\equiv\cos\theta_{ij}$, $\delta$ is the Dirac $C\!P$-violating phase, $\Phi$ is
a Majorana $C\!P$-violating phase. Neglecting statistical errors, we will use
throughout our study the values $\theta_{12}=\pi/5$ and $\theta_{23}=\pi/4$,
compatible with the results from neutrino oscillation experiments.
Moreover, we will adopt the $3\sigma$ range $s_{13}=0\textrm{--}0.20$.
\section{Baryogenesis via Leptogenesis: $M_{\nu}$ and Lower bound on $M_{\rho_3}$}
The leptogenesis mechanism was investigated in the context
of the Type III seesaw~\cite{TypeIII} mechanism in reference~\cite{Hambye}.
However, the case of hybrid seesaw, Type I plus Type III, has not been
investigated before, and it is the main focus of our work.
As emphasized in the previous section, the model predicts
$\rho_3$ lighter than $\rho_0$, so that, in a first approximation,
we will focus on the decay of $\rho_3$ to generate the observed
baryon asymmetry of the Universe via leptogenesis.

One of Sakharov's necessary conditions~\cite{sakharov} to satisfy in order to
produce a baryon asymmetry in the early Universe is $C\!P$ violation, which naturally
occurs in our model in the decays of $\rho_3$. We define the $C\!P$ asymmetry parameter as
\begin{equation}
\varepsilon_{\rho_3,\alpha}=-{\Gamma(\rho_3\to \ell_{\alpha}H^{\dagger})-
\Gamma(\rho_3\to \bar{\ell}_{\alpha} H)
\over \sum_{\alpha}\Gamma(\rho_3\to \ell_{\alpha}H^{\dagger})+\Gamma
(\rho_3\to \bar{\ell}_{\alpha} H)}.
\end{equation}
In the pure Type III case, the $C\!P$ asymmetry was computed in \cite{Hambye}, and was
found to differ from the pure Type I case by a factor 1/3 in the very hierarchical
limit. Here, the computation is slightly different from \cite{Hambye} because we have
one singlet running in the loops of the self-energy and the vertex corrections. The group
theory product is therefore different. Interestingly, the self-energy contribution
vanishes in our case. Hence, the only non-vanishing contribution is the vertex correction,
which turns out to have the same magnitude and sign as in the Type I seesaw. The
$C\!P$ asymmetry in our model is therefore given by~\cite{crv}
\begin{equation}\label{CPasym}
\varepsilon_{\rho_3,\alpha}=\frac{1}{8 \pi (h^{\dagger}h)_{11}}
{\rm Im}\left[h_{\alpha 1}^{\star}
h_{\alpha 2}(h^{\dagger}h)_{1 2}\right] f(M_{\rho_0}^2/M_{\rho_3}^2),
\end{equation}
where
\begin{equation}\label{xi}
f(x)= \,\sqrt{x}\,
\left[(1+x)\,\ln\left({1+x\over x}\right)-1\right] \, \stackrel{x\gg 1}{\to}
{1\over 2\sqrt{x}}\, ,
\end{equation}
which is a factor 3 smaller than in the Type I case in the hierarchical
limit $M_{\rho_3}\ll M_{\rho_0}$. It should be noted that, even though the $C\!P$ asymmetry
in our model has the same magnitude in the hierarchical limit as in the Type III seesaw, it
is for a completely different reason. In the latter case, the factor 1/3 shows up
because the vertex correction, which takes a negative sign, partially cancels the
self-energy contribution, which does not vanish in that case~\cite{Hambye}.

Note that the loop factor $f(x)$ in the $C\!P$ asymmetry from the singlet 
$\rho_0$ with the triplet $\rho_3$ running in the loop
is suppressed by a factor $M_{\rho_3}/M_{\rho_0}$~\cite{dibari}, which is required by 
the unification constraints to be less than $1/40$~\cite{Hoernisa}. Therefore,
without even considering washout aspects, 
the contribution to leptogenesis by the singlet can be safely neglected.

The third Sakharov's condition, namely departure from thermal equilibrium,
can be conveniently described by the so-called decay parameter
$K\equiv \widetilde{\Gamma}/H_{T=M_{\rho_3}}$, given by the ratio of the decay
widths to the expansion rate when $\rho_3$ starts to become non-relativistic
at $T=M_{\rho_3}$. In terms of Yukawa couplings, the decay parameters can be written as
\begin{equation}
\label{KalphaK}
K_{\alpha}={v_0^2 \over m_{\star} M_{\rho_3}}|h_{\alpha 1}|^2,\qquad \text{and} \qquad
\sum_{\alpha} K_{\alpha}=K={v_0^2 \over m_{\star} M_{\rho_3}}(h^{\dagger}h)_{11},
\end{equation}
where $m_{\star}$ is the equilibrium neutrino mass \cite{annals} given by
\begin{equation}\label{d}
m_{\star} \simeq 1.08\times 10^{-3}\,{\rm eV}.
\end{equation}
Let us now express the quantities in Eqs.~(\ref{CPasym}) and (\ref{KalphaK}) in terms
of the convenient Casas-Ibarra parametrization introduced above, Eq.~(\ref{ci}).
In the case of normal hierarchy, the flavored decay
parameters and their sum can be written respectively as
\begin{eqnarray}\label{param1}
K_{\alpha}&=&{m_{2}|U_{\alpha 2}|^2 \over m_{\star}}|1-\omega^2 | +
{m_3|U_{\alpha 3}|^2 \over m_{\star}}|\omega^2 |
\pm \xi {2\sqrt{m_2 m_3}
\over m_{\star}} {\rm Re} \left(U_{\alpha2}U_{\alpha3}^{\star}
\sqrt{1-\omega^2 }\omega ^{\star}\right),\\\label{param2}
K&=&{m_{2}\over m_{\star}}|1-\omega^2 | +
{m_3 \over m_{\star}}|\omega^2 |,
\end{eqnarray}
and since the mass splitting in the model is quite large,
as pointed out above, the flavored CP asymmetries read as
\begin{eqnarray}\label{param3}
\varepsilon_{\rho_3,\alpha}&\simeq& -{1\over 8 \pi v_0^2}{M_{\rho_3}\over m_2|1-\omega ^2|
+m_3 |\omega |^2}\left[ \left(m_3^2 |U_{\alpha 3}|^2-m_2^2 |U_{\alpha 2}|^2\right)
{\rm Im} (\omega ^2)\right. \nonumber\\
&&\pm \xi \sqrt{m_2 m_3}(m_3+m_2)\,{\rm Re} (U_{\alpha 2}^{\star} U_{\alpha 3})\,
{\rm Im} \left(\omega \sqrt{1-\omega ^2}\right)\nonumber \\
&&\left.\pm \xi \sqrt{m_2 m_3}(m_3-m_2)\,{\rm Im}( U_{\alpha 2}^{\star} U_{\alpha 3})\,
{\rm Re} \left(\omega \sqrt{1-\omega ^2}\right)\right].\label{flavCP}
\end{eqnarray}
The total $C\!P$ asymmetry $\varepsilon_{\rho_3}=\sum_{\alpha}\varepsilon_{\rho_3,\alpha}$
can be readily obtained from the latter expression:
\begin{equation}\label{param4}
\varepsilon_{\rho_3}\simeq -{1\over 8 \pi v_0^2}{M_{\rho_3}\over m_2|1-\omega ^2|
+m_3 |\omega |^2} \left(m_3^2 -m_2^2 \right)
{\rm Im} (\omega ^2).
\end{equation}
Note that the case of inverted hierarchy in Eqs.~(\ref{param1})--(\ref{param4})
is obtained by changing the indices $3\to 2$
and $2\to 1$. It can be noticed from the two above expressions that
the factor $\xi$ does not open any new region in the parameter space since
it always multiplies $\pm 1$. We will therefore assume $\xi=1$ in the following.

Now that we have defined the essential quantities for leptogenesis,
we can turn to the Boltzmann equations, which will have to be written in two
different regimes, the two-flavor regime and the unflavored regime. As a matter
of fact, when the mass range for $\rho_3$ is between $10^{9}$~GeV and
$5\times 10^{11}$~GeV, flavor effects cannot be neglected, and
the so-called two-flavour regime applies, with flavors denoted
$\alpha=e\mu, \tau$ \cite{nardi,abada,zeno}. For the range of
masses  $5\times 10^{11} \textrm{--} 10^{15}~{\rm GeV}$ we will
use unflavored Boltzmann equations.
\subsection{Flavored Regime }
Let us first discuss the two-flavor regime, which will give the
lowest bound on $M_{\rho_3}$. The relevant Boltzmann equations,
taking into account decays and inverse decays with proper
subtraction of the resonant contribution from $\Delta L=2$~\cite{giudice} and
$\Delta L=0$~\cite{pilaund,nardi} processes as well as $\Delta L=1$
scatterings~\cite{luty,abada2}, are given by
\begin{eqnarray}\label{flke1}
{dN_{\rho_3}\over dz} & = & -(D+S)\,(N_{\rho_3}-N_{\rho_3}^{\rm eq})
-2\, S_g(N_{\rho_3}^2-(N_{\rho_3}^{\rm eq})^2)\, , \\\label{flke2}
{dN_{\Delta_{\alpha}}\over dz} & = &
\varepsilon_{\rho_3,\alpha}(D+S)(N_{\rho_3}-N_{\rho_3}^{\rm eq})
- W^{\rm ID}_{\alpha}\sum_{\beta} C_{\alpha \beta} N_{\Delta_{\beta}} -
W^{\Delta L=1}_{\alpha}\sum_{\beta} C'_{\alpha \beta} N_{\Delta_{\beta}} \, ,
\end{eqnarray}
where $z \equiv M_{\rho_3}/T$, $\Delta_{\alpha}=B/3 -L_{\alpha}$,
and we indicated with $N_X$ any particle number or asymmetry $X$
calculated in a portion of co-moving volume containing one
$\rho_3$ (i.e. three components) in ultra-relativistic thermal
equilibrium, so that $N^{\rm eq}_{\rho_3}(T\gg M_{\rho_3})=1$.
The decay factor is given by
\begin{equation}
D \equiv {\Gamma_{D}\over H\,z}=3 K \,z\,
\left\langle {1\over\gamma} \right\rangle   \, ,
\end{equation}
where $H$ is the expansion rate and the factor 3 comes again from
the three components of $\rho_3$. The total decay rate,
$\Gamma_{D}\equiv \Gamma+\bar{\Gamma}$,
is the product of the decay width times the
thermally averaged dilation factor $\langle 1/\gamma\rangle$, given by the ratio
${\cal K}_1(z)/ {\cal K}_2(z)$ of the modified Bessel functions.
A simple analytic approximation for the sum $D+S$, where $S$ is the
contribution from the Higgs-mediated
scattering processes, is given by~\cite{annals}
\begin{equation}
D+S\simeq 3\times 0.1\, K\left[1+\ln\left({M_{\rho_3}\over M_h}\right)z^2
\ln\left(1+{a\over z}\right)\right],
\end{equation}
where $M_h$ is the Higgs mass and
\begin{equation}
a={8\pi^2\over 9 \ln(M_{\rho_3}/M_h)}.
\end{equation}
The equilibrium abundance and its rate are
also expressed through the modified Bessel functions,
\begin{equation}
N_{\rho_3}^{\rm eq}(z)= {1\over 2}\,z^2\,{\cal K}_2 (z) \;\; ,
\qquad \text{and} \qquad
{dN_{\rho_3}^{\rm eq}\over dz} =
-{1\over 2}\,z^2\,{\cal K}_1 (z) \, .
\end{equation}
The inverse decay washout term, with the resonant $\Delta L=2$
contribution properly subtracted~\cite{giudice}, is given by
\begin{equation}\label{WID}
W^{\rm ID}_{\alpha}(z)={3\over 4}\,K_{\alpha}\ {\cal K}_1(z)\,z^3.
\end{equation}
It was shown in~\cite{annals} that the complete washout term can
be conveniently expressed as
\begin{equation}
W^{\rm ID}_{\alpha}(z)+W^{\Delta L=1}_{\alpha} = j(z) W^{\rm ID}_{\alpha}(z),
\end{equation}
where
\begin{equation}
j(z)=0.1 \left(1+{15\over 8 z}\right)\left[z\ln\left({M_{\rho_3}\over M_h}\right)
\ln \left(1+{a\over z}\right)+{1\over z}\right]
\end{equation}
in the strong washout regime, which will be the relevant one in the subsequent
discussion.

The effects of the so-called `spectator processes' \cite{buchplum,nardi2}, which
translate into a non-trivial relation between the asymmetries stored in the
lepton doublets $\ell_{\alpha}$ and the asymmetries $\Delta_{\alpha}$ \cite{bcst}
as well as into an additional washout due to the asymmetry in the Higgs
field~\cite{nardi2}, are accounted for by the matrices
$C$~\cite{newaspects,nardi2} and $C'$, whose components are given by
\begin{equation}\label{C}
C = \left(\begin{array}{cc}
1.11 & 0.25\\
0.21 & 1.01\end{array}\right),\qquad \text{and} \qquad C' = \left(\begin{array}{cc}
0.98 & 0.08\\
0.08 & 0.84\end{array}\right) .
\end{equation}
The matrix $C'$ is different from $C$ because the Higgs asymmetry contribution
is divided by 2 in the $\Delta L=1$ scattering case~\cite{nardi2}.

Compared to the `usual' computation with singlet neutrinos, there is a
new term in the equation for the abundance of $\rho_3$, as pointed out
in \cite{Hambye}. It originates from scatterings allowed by the interaction
Eq.~(\ref{scattint}). From the calculation in \cite{Hambye} we found the
useful fit (within 30\% accuracy in the relevant range $0.1\lesssim z \lesssim 10$)
\begin{equation}
S_g\simeq 10^{-3} {M_{\rm Pl}\over M_{\rho_3}} {\sqrt{1+{\pi\over 2}z^{-0.3}}\over
(15/8+z)^2(1+{\pi z/ 2})} e^{0.3 z}.
\end{equation}
The small uncertainty introduced by using this fit
will translate into less than 10\% effects on the final baryon asymmetry.

In writing Eqs.~(\ref{flke1})--(\ref{flke2}) we are neglecting
non-resonant $\Delta L=2$ processes contribution
and $\Delta L=0$ processes, a good approximation for
$M_1\ll 10^{14}\,{\rm GeV}\,(m_{\rm atm}^2/\sum_i\,m_i^2)$, certainly satisfied
in the flavored regime. We are also neglecting thermal corrections~\cite{giudice},
which are expected to be small in the strong washout regime.

Solving the Boltzmann equations (\ref{flke1})--(\ref{flke2}), one obtains
$N_{\Delta_{\alpha}}^{\rm f}=N_{\Delta_{\alpha}}(z\to \infty)$, and hence
the baryon-to-photon ratio predicted is
\begin{equation}
\eta_B\simeq 3\times0.96\times 10^{-2} \left(N_{\Delta_{e\mu}}^{\rm f}+N_{\Delta_{\tau}}^{\rm f}\right),
\end{equation}
where the factor 3 comes from the three degrees of freedom in the
fermionic triplet $\rho_3$. This prediction must then be compared with the
observed value~\cite{WMAP5}
\begin{equation}\label{etaBobs}
\eta_B^{\rm CMB} = (6.2 \pm 0.15)\times 10^{-10} \, .
\end{equation}
%
\subsection{Unflavored Regime}
Let us now discuss the unflavored regime, for $5\times 10^{11}~{\rm GeV}<M_{\rho_3}
<10^{15}~{\rm GeV}$. The relevant Boltzmann equations are
\begin{eqnarray}\label{unflke1}
{dN_{\rho_3}\over dz} & = & -(D+S)\,(N_{\rho_3}-N_{\rho_3}^{\rm eq})
-2\, S_g \,(N_{\rho_3}^2-(N_{\rho_3}^{\rm eq})^2)\, ,\\\label{unflke2}
{dN_{B-L}\over dz} & = &
\varepsilon_{\rho_3}(D+S)(N_{\rho_3}-N_{\rho_3}^{\rm eq})
- W N_{B-L} ,
\end{eqnarray}
where $\varepsilon_{\rho_3}=\sum_{\alpha}\varepsilon_{\rho_3,\alpha}$ and $W(z)=j(z)\sum_{\alpha}
 W^{\rm ID}_{\alpha}(z)+
\Delta W(z)$. The contribution to the washout by the non-resonant
$\Delta L=2$ processes, $\Delta W(z)$, is given by~\cite{annals}
\begin{equation}
\Delta W(z)\simeq 3\times 10^{-3}\,{0.186\over z^2}\,\left({M_{\rho_3}\over 10^{10}~{\rm GeV}}\right)
\left({\overline{m}^2\over {\rm eV}^2}\right),
\end{equation}
with $\overline{m}^2\equiv m_1^2+m_2^2+m_3^2= 2.7\,(4.9)~{\rm eV}^2$ for normal
(inverted) hierarchy.

After solving Eqs.~(\ref{unflke1})--(\ref{unflke2}), one obtains
$N_{B-L}^{\rm f}=N_{B-L}(z\to \infty)$, from which the final baryon asymmetry
\begin{equation}
\eta_B\simeq 3\times0.96\times 10^{-2}\,N_{B-L}^{\rm f} ,
\end{equation}
is derived, to be compared with the measured value, Eq.~(\ref{etaBobs}).
\subsection{Numerical Results}
\begin{figure}
\begin{center}
\includegraphics[width=0.6\textwidth, angle=-90]{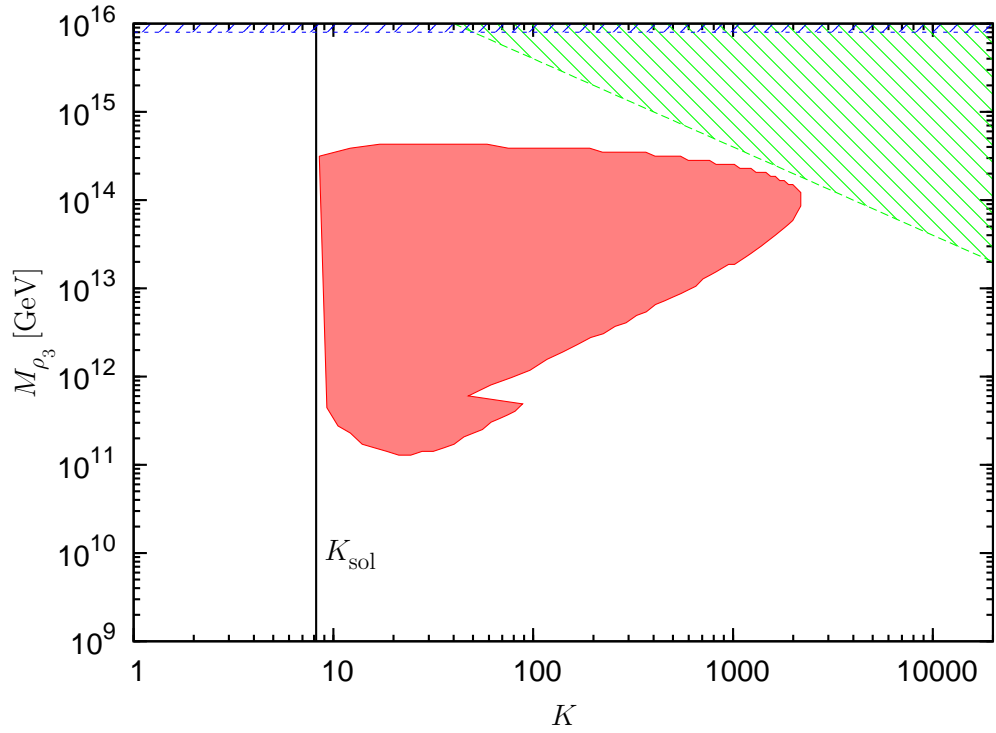}
\caption{Allowed range for the mass of $\rho_3$ vs. the decay parameter
$K$ (pink region). Case of normal hierarchy.}\label{fig:SMnormal}
\end{center}
\end{figure}

\begin{figure}
\begin{center}
\includegraphics[width=0.6\textwidth, angle=-90]{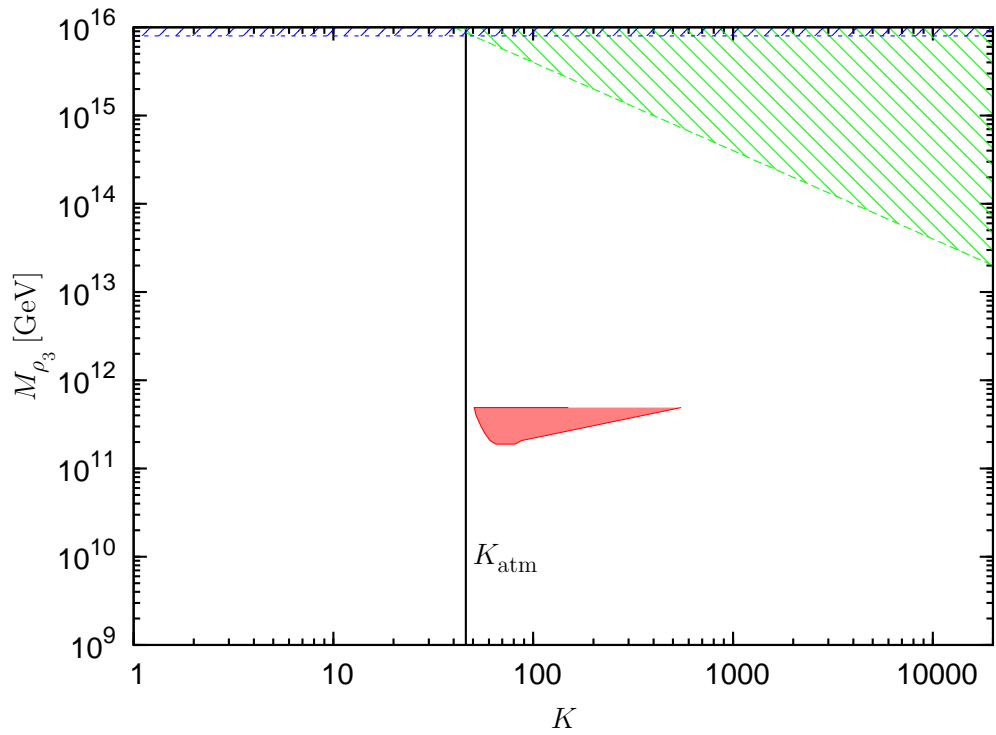}
\caption{Allowed range for the mass of $\rho_3$ vs. the decay parameter
$K$ (pink region). Case of inverted hierarchy.}\label{fig:SMinverted}
\end{center}
\end{figure}

In order to obtain the region in the parameter space $(K, M_{\rho_3})$ that
is allowed by successful leptogenesis, one needs to solve the Boltzmann
equations in the two-flavor regime,
Eqs.~(\ref{flke1})--(\ref{flke2}), and then to maximize the asymmetry
over all unknown parameters $(\theta_{13}, \delta,
\Phi, \omega )$ at every given value of $K$. In the unflavored regime, one
needs to solve Eqs.~(\ref{unflke1})--(\ref{unflke2}) and maximize
over $\omega$ at every given value of $K$. The result is shown in
Fig.~\ref{fig:SMnormal} for a normal hierarchy of light neutrinos and
in Fig.~\ref{fig:SMinverted} for an inverted hierarchy.

Let us explain the origin of the shaded areas in the plots. Imposing that the Yukawa
couplings $h_{\alpha 1,2}$ remain perturbative, i.e. $h_{\alpha 1,2}<2\sqrt{\pi}$,
implies
\begin{equation}
M_{\rho_3} \ < \ {4\times 10^{17}~{\rm GeV}\over K},
\end{equation}
excluding the triangle shaded area in the figures.
Additionally, the atmospheric scale must be accounted for, i.e.
$h_{\alpha 1,2}^2 v_0^2/M_{\rho_3}\gtrsim m_{\rm atm}$, implying
\begin{equation}
M_{\rho_3} \ < \ 8 \times 10^{15}~{\rm GeV},
\end{equation}
which excludes the horizontal shaded area in the figures.
Turning to Fig.~\ref{fig:SMnormal}, one can clearly see the transition from
the two-flavor to the unflavored regime, when the mass of $\rho_3$
goes over $5\times 10^{11}$~GeV. Flavor effects would introduce a relaxation
of the lowest bound by roughly an order of magnitude if the gauge scattering term $S_g$
was not present, but accounting for them, the relaxation is only by a factor 2-3. We
therefore confirm that the gauge scattering term induces a reduction of the maximal efficiency
factor, as pointed out in \cite{Hambye} and \cite{Fischler}. Furthermore,
we would like to point out that spectator processes, whose effects are accounted
for in $C$ and $C'$ in Eq.~(\ref{flke2}), induce a reduction of the allowed region
by about 30\% in the flavored regime. As for the $\Delta L=1$ scatterings, their
inclusion changes the final asymmetry only very marginally, confirming what was found
in~\cite{flavorlep}.

A nice feature of the computation is that the final asymmetry is insensitive to the initial
number of $\rho_3$. The reason is that one has a regime of strong washout,
also when flavor effects are included. The gauge scattering term, which quickly thermalizes
the abundance of $\rho_3$, also contributes to the independence
of the initial number of $\rho_3$. The strong washout is ensured by the fact
that $K\geq K_{\rm sol}\equiv 8.2\gg 1$ in the case of normal hierarchy and
$K\geq K_{\rm atm}\equiv 46\gg 1$ in the case of inverted hierarchy.

In Fig.~\ref{fig:SMinverted}, where the case of inverted hierarchy is displayed,
it is apparent that the allowed region is very small. Actually, only in the flavored
regime below $5\times 10^{11}$~GeV there is an allowed region. In the unflavored
regime the usual suppression of the $C\!P$ asymmetry for the case of two right-handed neutrinos in the
inverted hierarchy \cite{blanchet}, combined with the washout from the non-resonant $\Delta L=2$ processes, leaves
no allowed region. On the other hand, when flavor effects are included, the $C\!P$
asymmetry is not suppressed, and the final asymmetry can be orders of magnitude higher
than what would be predicted with an unflavored calculation. This was already noticed
in the very similar case of two singlet neutrinos in~\cite{abada2} and confirmed
recently in~\cite{newaspects}. It is important to say that there is a big sensitivity
to the `low-energy' $C\!P$-violating phases in the PMNS matrix in that case which will 
be studied in a future publication. This behavior in our model is not 
surprising if one remembers the similarity
with the 2 right-handed neutrino model, where the crucial role by the $C\!P$-violating 
phases in the PMNS matrix in the case of inverted hierarchy was recently 
emphasized in~\cite{newaspects} (see the right panel of Fig.~5).

\section{Constraints from Leptogenesis and the Spectrum of the Theory}
\begin{figure}
\begin{center}
\includegraphics[width=0.6\textwidth, angle=0]{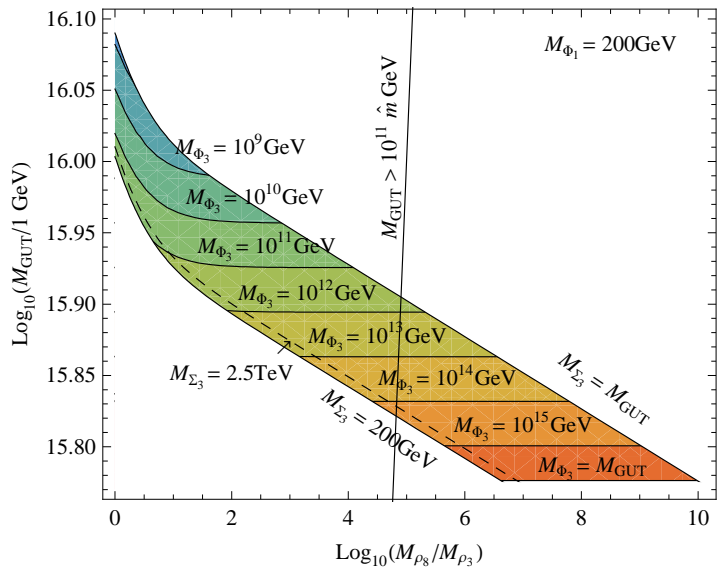}
\caption{Constraints coming from unification when $M_{\Phi_1}=200$ GeV,
including the bound from leptogenesis.}
\label{plot1}
\end{center}
\end{figure}

In the previous section we have obtained the allowed region by
leptogenesis for each neutrino mass spectrum. In the case of Normal Hierarchy
we find
\begin{equation}
 10^{11.1} \ \text{GeV} \lesssim M_{\rho_3}^{\rm NH} \lesssim 10^{14.5} \ \text{GeV}\, ,
\end{equation}
while for Inverted Hierarchy we find
\begin{equation}
10^{11.3} \ \text{GeV} \lesssim M_{\rho_3}^{\rm IH} \lesssim 10^{11.7} \ \text{GeV}\, .
\end{equation}
In reference~\cite{Hoernisa} the solutions
for the spectrum in the theory which are consistent with the unification of gauge
interactions and proton decay were shown. Let us see what is the role of the above
constraints coming from leptogenesis. The relation between the masses of the
fermionic fields living in the adjoint representation is given by
\begin{equation}
M_{\rho_0}= \frac{1}{5} \left( 3 + 2 \hat{m}\right)\, M_{\rho_3}, \
\quad \ M_{\rho_8}=\hat{m} M_{\rho_3},
\ \quad \ \text{and} \ \quad \ M_{\rho_{(3,2)}}=M_{\rho_{(\bar{3},2)}}=
\frac{1}{2} \left( 1 + \hat{m} \right) M_{\rho_3}.
\end{equation}
Since the mass of all these fields should be below the GUT scale, one can use these relations
as well as
the bounds on $M_{\rho_3}$ coming from leptogenesis in order to constrain the spectrum. The
most relevant constraint comes from the relation between $M_{\rho_8}$ and
$M_{\rho_3}$, from which we find
\begin{equation}\label{mhat}
 10^{11.1} \ \text{GeV} \ \hat{m} \lesssim M_{\rm GUT} \lesssim 10^{15.90} \ \text{GeV},
\end{equation}
where the upper bound is coming from the unification and proton decay
constraints~\cite{Hoernisa} and the lower bound is due to the minimal allowed
value for $M_{\rho_3}$. We recall that $\hat{m}$ is the mass splitting.
From Eq.~(\ref{mhat}) one readily obtains an upper bound on the mass splitting
between the fields in the adjoint representation:
\begin{equation}
10^{2} \lesssim \hat{m} \lesssim 10^{4.8},
\end{equation}
where the lower bound comes from the unification constraints~\cite{Hoernisa}.
Therefore, one excludes a large part of the allowed parameter space shown
in reference~\cite{Hoernisa}.  In order to show the importance of these
bounds we present in Fig.~\ref{plot1} the parameter
space allowed by
unification when $M_{\Phi_1}=200$ GeV. The fields $\Phi_1 \sim (8,2,1/2)$
and $\Phi_3 \sim (3,3,-1/3)$ live in ${\bf 45_H}$ while
$\Sigma_3 \sim (1,3,0)$ is in ${\bf 24_H}$. See reference~\cite{Hoernisa}
for details.
Notice that once we include the leptogenesis constraints a large part
of the parameter space is excluded. Now, since $M_{\Phi_3}$ has to be larger
than $10^{12}$ GeV in order to satisfy the constraints coming from
proton decay, the only allowed region in Fig.~\ref{plot1} is the area bounded
by the lines $M_{\Sigma_3} = 200$ GeV, $M_{\Phi_3}= 10^{12}$ GeV
and $M_{\rm GUT} = 10^{11} \hat{m}$ GeV. This means that the model
is quite predictive.

\section{Summary}
We have presented a detailed study of the Baryogenesis via Leptogenesis
mechanism in the context of Adjoint $SU(5)$ where neutrino masses
are generated by the Type I and Type III seesaw mechanisms. Through
the decays of the field responsible for the Type III seesaw, $\rho_3$,
a lepton asymmetry is produced and later converted in a baryon asymmetry
by sphalerons. In our model, the $C\!P$ asymmetry is generated by the
vertex correction since the self-energy contribution vanishes. Imposing
successful leptogenesis, we found that the case of normal hierarchy for
the neutrinos is possible for a large range of $\rho_3$ masses 
(see Fig.~\ref{fig:SMnormal}). On the other hand,
when the spectrum is inverted, the allowed region is very small
(see Fig.~\ref{fig:SMinverted}).
Finally, we have shown that, imposing successful leptogenesis, one rules
out a large region in the parameter space allowed by the unification
of gauge interactions and the constraints coming from
proton decay (see Fig.~\ref{plot1}).
\begin{acknowledgments}
We would like to thank Pasquale Di Bari for discussions and comments 
on the manuscript.
The work of P. F. P. was supported in part by the U.S. Department of Energy
contract No. DE-FG02-08ER41531 and in part by the Wisconsin Alumni
Research Foundation.
\end{acknowledgments}



\begin{thebibliography}{99}

\bibitem{Fukugita}
  M.~Fukugita and T.~Yanagida,
  ``Baryogenesis Without Grand Unification,''
  Phys.\ Lett.\  B {\bf 174} (1986) 45.

\bibitem{TypeI}
  P.~Minkowski,
  ``Mu $\to$ E Gamma At A Rate Of One Out Of 1-Billion Muon Decays?,''
  Phys.\ Lett.\ B {\bf 67} (1977) 421 ;
  T. Yanagida, in {\it Proceedings of the Workshop on the Unified Theory
   and the Baryon Number in the Universe}, eds. O. Sawada et al., (KEK
   Report~79-18, Tsukuba, 1979), p.~95;
  M. Gell-Mann, P. Ramond and R. Slansky,
   in {\it Supergravity}, eds. P. van Nieuwenhuizen et al.,
   (North-Holland, 1979), p.~315;
  S.L. Glashow, in {\it Quarks and Leptons}, Carg\`ese, eds. M. L\'evy et al.,
(Plenum, 1980), p. 707;
  R.~N.~Mohapatra and G.~Senjanovi\'c,
  ``Neutrino Mass And Spontaneous Parity Nonconservation,''
  Phys.\ Rev.\ Lett.\  {\bf 44} (1980) 912.

\bibitem{kuzmin}
  V.~A.~Kuzmin, V.~A.~Rubakov and M.~E.~Shaposhnikov,
  ``On The Anomalous Electroweak Baryon Number Nonconservation In The Early
  Universe,''
  Phys.\ Lett.\  B {\bf 155} (1985) 36.

\bibitem{GUTs}
  For a recent review see: P.~Nath and P.~Fileviez P\'erez,
  ``Proton stability in grand unified theories, in strings, and in branes,''
  Phys.\ Rept.\  {\bf 441} (2007) 191
  [arXiv:hep-ph/0601023].


\bibitem{adjoint}
  P.~Fileviez~P\'erez,
  ``Renormalizable Adjoint SU(5),''
  Phys.\ Lett.\  B {\bf 654} (2007) 189
  [arXiv:hep-ph/0702287].

\bibitem{TypeIII}
  R.~Foot, H.~Lew, X.~G.~He and G.~C.~Joshi,
  ``Seesaw neutrino masses induced by a triplet of leptons,''
  Z.\ Phys.\ C {\bf 44} (1989) 441.

\bibitem{Hoernisa}
  P.~Fileviez P\'erez, H.~Iminniyaz and G.~Rodrigo,
  ``Proton Stability, Dark Matter and Light Color Octet Scalars in Adjoint
  SU(5) Unification,'' Phys.\ Rev.\  D {\bf 78} (2008) 015013
  arXiv:0803.4156 [hep-ph].

\bibitem{SUSY}
  P.~Fileviez~P\'erez,
  ``Supersymmetric Adjoint SU(5),''
  Phys.\ Rev.\  D {\bf 76} (2007) 071701
  [arXiv:0705.3589 [hep-ph]].


\bibitem{review}
  M.~C.~Gonzalez-Garcia and M.~Maltoni,
  ``Phenomenology with Massive Neutrinos,''
  Phys.\ Rept.\  {\bf 460} (2008) 1
  [arXiv:0704.1800 [hep-ph]];
  A.~Strumia and F.~Vissani,
  ``Neutrino masses and mixings and.,''
  arXiv:hep-ph/0606054;
  R.~N.~Mohapatra and A.~Y.~Smirnov,
  ``Neutrino mass and new physics,''
  Ann.\ Rev.\ Nucl.\ Part.\ Sci.\  {\bf 56} (2006) 569
  [arXiv:hep-ph/0603118];
  V.~Barger, D.~Marfatia and K.~Whisnant,
  ``Progress in the physics of massive neutrinos,''
  Int.\ J.\ Mod.\ Phys.\  E {\bf 12} (2003) 569
  [arXiv:hep-ph/0308123].

\bibitem{casas}
J.~A.~Casas and A.~Ibarra,
``Oscillating neutrinos and mu $\to$ e, gamma,''
Nucl.\ Phys.\ B {\bf 618} (2001) 171.

\bibitem{ir}
  A.~Ibarra and G.~G.~Ross,
  ``Neutrino properties from Yukawa structure,''
  Phys.\ Lett.\  B {\bf 575}, 279 (2003)
  [arXiv:hep-ph/0307051];
 A.~Ibarra and G.~G.~Ross,
  ``Neutrino phenomenology: The case of two right handed neutrinos,''
  Phys.\ Lett.\  B {\bf 591}, 285 (2004)
  [arXiv:hep-ph/0312138].

\bibitem{PDG}
  S.~Eidelman et al., Phys.\ Lett. {\bf B592}, 1 (2004)
  (URL:http://pdg.lbl.gov/).

\bibitem{Hambye}
  T.~Hambye, Y.~Lin, A.~Notari, M.~Papucci and A.~Strumia,
  ``Constraints on neutrino masses from leptogenesis models,''
  Nucl.\ Phys.\  B {\bf 695} (2004) 169
  [arXiv:hep-ph/0312203];

  A.~Strumia,
  ``Sommerfeld corrections to type-II and III leptogenesis,''
  arXiv:0806.1630 [hep-ph].

\bibitem{sakharov}
  A.~D.~Sakharov,
  ``Violation of CP Invariance, c Asymmetry, and Baryon Asymmetry of the
  Universe,''
  Pisma Zh.\ Eksp.\ Teor.\ Fiz.\  {\bf 5} (1967) 32
  [JETP Lett.\  {\bf 5} (1967\ SOPUA,34,392-393.1991\ UFNAA,161,61-64.1991) 24].

\bibitem{crv}
  L.~Covi, E.~Roulet and F.~Vissani,
  ``CP violating decays in leptogenesis scenarios,''
  Phys.\ Lett.\  B {\bf 384} (1996) 169
  [arXiv:hep-ph/9605319].

\bibitem{dibari}
  P.~Di Bari,
  ``Seesaw geometry and leptogenesis,''
  Nucl.\ Phys.\  B {\bf 727} (2005) 318
  [arXiv:hep-ph/0502082].

\bibitem{annals}
  W.~Buchmuller, P.~Di Bari and M.~Plumacher,
  ``Leptogenesis for pedestrians,''
  Annals Phys.\  {\bf 315} (2005) 305
  [arXiv:hep-ph/0401240].

\bibitem{nardi}
  E.~Nardi, Y.~Nir, E.~Roulet and J.~Racker,
  ``The importance of flavor in leptogenesis,''
  JHEP {\bf 0601} (2006) 164
  [arXiv:hep-ph/0601084].

\bibitem{abada}
  A.~Abada, S.~Davidson, F.~X.~Josse-Michaux, M.~Losada and A.~Riotto,
  ``Flavour issues in leptogenesis,''
  JCAP {\bf 0604} (2006) 004
  [arXiv:hep-ph/0601083].


\bibitem{zeno}
  S.~Blanchet, P.~Di Bari and G.~G.~Raffelt,
  ``Quantum Zeno effect and the impact of flavor in leptogenesis,''
  JCAP {\bf 0703} (2007) 012
  [arXiv:hep-ph/0611337].



\bibitem{giudice}
  G.~F.~Giudice, A.~Notari, M.~Raidal, A.~Riotto and A.~Strumia,
  ``Towards a complete theory of thermal leptogenesis in the SM and MSSM,''
  Nucl.\ Phys.\  B {\bf 685} (2004) 89
  [arXiv:hep-ph/0310123].



\bibitem{pilaund}
  A.~Pilaftsis and T.~E.~J.~Underwood,
 ``Electroweak-scale resonant leptogenesis,''
  Phys.\ Rev.\  D {\bf 72} (2005) 113001
  [arXiv:hep-ph/0506107].

\bibitem{luty}
  M.~A.~Luty,
  ``Baryogenesis Via Leptogenesis,''
  Phys.\ Rev.\  D {\bf 45} (1992) 455.

\bibitem{abada2}
  A.~Abada, S.~Davidson, A.~Ibarra, F.~X.~Josse-Michaux, M.~Losada and A.~Riotto,
  ``Flavour matters in leptogenesis,''
  JHEP {\bf 0609} (2006) 010
  [arXiv:hep-ph/0605281].

\bibitem{buchplum}
  W.~Buchm\"{u}ller and M.~Pl\"{u}macher,
  ``Spectator processes and baryogenesis,''
  Phys.\ Lett.\ B {\bf 511} (2001) 74 [arXiv:hep-ph/0104189].

\bibitem{nardi2}
  E.~Nardi, Y.~Nir, J.~Racker and E.~Roulet,
  ``On Higgs and sphaleron effects during the leptogenesis era,''
  JHEP {\bf 0601} (2006) 068
  [arXiv:hep-ph/0512052].

\bibitem{bcst}
  R.~Barbieri, P.~Creminelli, A.~Strumia and N.~Tetradis,
  ``Baryogenesis through leptogenesis,''
  Nucl.\ Phys.\  B {\bf 575} (2000) 61
  [arXiv:hep-ph/9911315].

\bibitem{newaspects}
  S.~Blanchet and P.~Di Bari,
  ``New aspects of leptogenesis bounds,''
  arXiv:0807.0743 [hep-ph].

\bibitem{WMAP5}
  E.~Komatsu {\it et al.}  [WMAP Collaboration],
  ``Five-Year Wilkinson Microwave Anisotropy Probe (WMAP)
  Observations:Cosmological Interpretation,''
  arXiv:0803.0547 [astro-ph].

\bibitem{Fischler}
  W.~Fischler and R.~Flauger,
  ``Neutrino Masses, Leptogenesis, and Unification in the Absence of Low Energy
  Supersymmetry,''
  arXiv:0805.3000 [hep-ph].

\bibitem{flavorlep}
  S.~Blanchet and P.~Di Bari,
``Flavor effects on leptogenesis predictions,''
   JCAP {\bf 03} (2007) 018.

\bibitem{blanchet}
  S.~Blanchet and P.~Di Bari,
  ``Leptogenesis beyond the limit of hierarchical heavy neutrino masses,''
  JCAP {\bf 0606} (2006) 023
  [arXiv:hep-ph/0603107].


\end{thebibliography}
\end{document}